\documentclass{article}

\usepackage{arxiv}

\usepackage[utf8]{inputenc} 
\usepackage[T1]{fontenc}    
\usepackage{hyperref}       
\usepackage{url}            
\usepackage{booktabs}       
\usepackage{amsfonts}       
\usepackage{nicefrac}       
\usepackage{microtype}      
\usepackage{lipsum}		
\usepackage{graphicx}
\usepackage{natbib}
\usepackage{doi}
\usepackage{caption} 
\title{A Bayesian Surveillance Model to Track Variable Rainfall-Runoff Responses for Small Watersheds}

\author{ \href{https://orcid.org/0000-0001-9402-2865}{\includegraphics[scale=0.06]{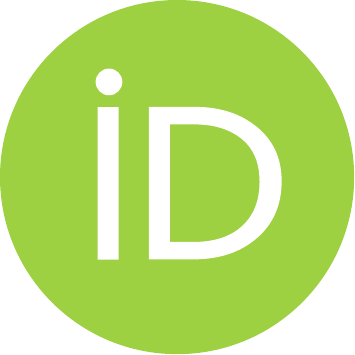}\hspace{1mm}Xiao Peng} \\
	School of Civil and Environmental Engineering\\
	Cornell University\\
	Ithaca, NY 14850 \\
	\texttt{xp53@cornell.edu} \\
	\And
	\href{https://orcid.org/0000-0002-7276-9789}{\includegraphics[scale=0.06]{orcid.pdf}\hspace{1mm}John D. Albertson} \thanks{Corresponding author}\\
	School of Civil and Environmental Engineering\\
	Cornell University\\
	Ithaca, NY 14850 \\
	\texttt{albertson@cornell.edu} \\
}

\renewcommand{\shorttitle}{Bayesian Surveillance for Rainfall-Runoff Responses}

\hypersetup{
pdftitle={MCMC-based diagnostic model for small watersheds},
pdfsubject={stat.AP},
pdfauthor={Xiao Peng, John D. Albertson},
pdfkeywords={MCMC, MAP, IUH},
}

\begin{document}
\maketitle

\begin{abstract}
Understanding dynamics of hydrological responses is essential in producing skillful runoff forecast. This can be quantitatively done by tracking changes in hydrology model parameters that represent physical characteristics. In this study, we implement a Bayesian estimation method in continuously estimating hydrology model parameters given observations of rainfall and runoff for small watersheds. The method is coupled with a conceptual hydrology model using a Gamma distribution-based Instantaneous Unit Hydrograph. The whole analytical framework is tested using synthetic data as well as observational data from the Fall Creek watershed. The results show that the Bayesian method can well track the hidden parameters that change inter-annually. Then the model is applied to examine temporal and spatial variability of the rainfall-runoff responses and we find 1) a systematic shift in the rainfall-runoff response for the Fall Creek watershed around 1943 and 2) a statistically significant relationship between rainfall-runoff responses and watershed sizes for selected NY watersheds. Our results demonstrate potential of the Bayesian estimation method as a rapid surveillance tool in monitoring and tracking changes of hydrological responses for small watersheds.

\end{abstract}


\section{Introduction}

Skillful runoff forecast is of great importance in informing decision making in water resource management across various sectors including but not limited to hydropower plant operation \citep{jiang2018runoff}, agriculture resource planning \citep{shah2017short}, and disaster mitigation \citep{pappenberger2005cascading,fundel2013monthly}. At long timescales (e.g., seasonal to inter-annual), runoff amplitudes are usually largely determined by rainfall amplitudes. But at short timescales (e.g., hourly to weekly), some key variables for defining high-frequency water disasters (e.g., flooding) like timings and rates of peak flow are also determined by the rainfall-runoff responses as controlled by the local hydrological properties \citep{cunderlik2009trends,do2020historical}. To understand dynamics of rainfall-runoff responses can aid the decision making of water managers by reducing uncertainties in the runoff forecasts/estimations \citep{montanari2004stochastic,moradkhani2005dual}.

The past decades have witness significant advances in sensing technology, which are making massive volumes of earth observation data available at exceptionally fine resolutions in space and time \citep{ma2015remote,chi2016big}. However, these observational data are highly concentrated in some easy-to-measure variables like rainfall and runoff. Though it is hard to directly measure the variable rainfall-runoff response or its hidden variables, such information can be inferred from observations of rainfall and runoff. Past studies already show that some parameters of physically-based hydrology models can reflect physical properties of watersheds like river morphology \citep{hassan2006experiments} or land use types \citep{heuvelmans2004evaluation}. These properties can determine the rainfall-runoff response and thus tracking estimations of the corresponding parameters can help us learn dynamics of the local responses \citep{nourani2017hydrological}. Based on this fact, the problem of learning non-stationary rainfall-runoff responses can be refined to estimating time-varying parameters of physics-informed hydrology models given observations of some easy-to-measure variables like rainfall and runoff.

Most methods designed for model calibration are suited for this job and they can be roughly categorized into two classes: 1) the Maximum Likelihood Estimation (MLE) method that searches for the optimal parameter values that maximize the likelihood function for a given set of observations \citep{sorooshian1983evaluation,castiglioni2010calibration}; and 2) the Bayesian method (also often referred to as the Maximum a Posteriori (MAP) method) that estimates the probability distribution function (PDF) of parameters given observations \citep{bates2001markov,smith2008bayesian}. Either method has its own advantages and represents different perspectives of approaching a problem. MLE is often described as the frequentists' methodology while MAP is often described as the Bayesian's methodology \citep{nolan2021frequentist}. For situations where information of prior distributions and sources of uncertainty are known, the Bayesian method is the preferable option since it can simultaneously estimate parameters and quantify the associated uncertainties \citep{sivia2006data}. In addition, for high-dimensional problems, the simulated PDF can converge to the true PDF at a controllable rate using Bayesian methods \citep{cowles1996markov} while the computational cost can increase exponentially as the number of parameters increases when using MLE methods (e.g., grid search methods) \citep{ba2017maximum}. These advantages make the Bayesian method suitable for inferring parameters of hydrology models here since most hydrology models use not multiple but many parameters \citep{devia2015review}. 

Bayesian estimation models has already been widely used in calibrating hydrology models. Some early effort can date back to 1970s when \citet{vicens1975bayesian} attempted to reduce parameter uncertainties by using informative prior PDFs based on regional information. This work focused on fusing information from prior knowledge and sample observations and resembles a lot of studies in data assimilation and filtering (e.g., the famous Kalman filter by \citet{kalman1960new}). Since then, the Bayesian estimation method has gained rapidly increasing popularity in uncertainty quantification for hydrology model parameters. \citet{yeh1986review} described the parameter identification procedures in groundwater hydrology as an inverse problem and introduced the Bayesian solution using examples of the composite objective function \citep{gavalas1976reservoir} and the Kalman filter. \citet{kitanidis1986parameter} compared information fusion processes in a Bayesian analysis with different prior and likelihood distributions (e.g., when the normality assumption is not satisfied). Further sophisticated Bayesian methods are applied in tuning hydrology models. \citet{bates2001markov} tested a Markov Chain-Monte Carlo scheme in inferring parameters for a conceptual rainfall-runoff model. \citet{moradkhani2005dual} implemented the ensemble Kalman filter in iterative hierarchical scheme to simultaneous estimate hyperparameters and state variables given observations of rainfall and runoff. Thanks to the rapid development of computation technology, a lot more simulation-based techniques are developed and tested in learning hydrology model parameters like the particle filter and its derivatives \citep{moradkhani2005uncertainty,manoli2015iterative,abbaszadeh2018enhancing}. While all these previous studies have validated the Bayesian estimation method in calibrating hydrology models, most of them are done in a context of data assimilation or uncertainty quantification. Even when \citet{moradkhani2005dual} managed to estimate time-varying parameters and the associated uncertainties, they highlighted their method as a potential tool to update hydrology simulations and decompose uncertainties from different sources. Here, we want to address the opportunity of using Bayesian estimation methods in learning dynamics of rainfall-runoff responses. And this can be quantitatively done by iteratively feeding the analysis with batches of rainfall and runoff observations collected from different historical periods or different watersheds. 

In this paper, we propose to test an analytical framework for monitoring changes in hydrological responses by tracking estimations of hydrology model parameters given rainfall and runoff observations. The framework is based on a Bayesian estimation method and is specifically designed for small watersheds. For large watersheds, a lot of information would be lost by aggregating spatially highly-heterogeneous hydrological properties into a few parameters and this would lead to huge uncertainties in parameter estimations. The model is tested using both synthetic and observational data of rainfall and runoff, and is applied to investigate temporal and spatial variability of rainfall-runoff responses.

\section{Data}

The analytical framework is first validated using synthetic rainfall and runoff data generated from historical records of the Fall Creek watershed. Fall Creek is a fourth-order stream that flows through Ithaca, NY and drains to Cayuga Lake. The watershed of Fall Creek has a relatively small drainage area of 324 km$^2$ and consists of mixed forested and agricultural areas \citep{knighton2017hydrologic}. Like most other NY watersheds, the Fall Creek watershed is dominated by a saturation-excess process and runoff is usually only generated when and where soils are highly saturated \citep{easton2007hydrologic,dahlke2009modelling}. The watershed is chosen for having relatively long records of ground-based precipitation and runoff observations at a daily timescale. Runoff data is collected from the National Water Information System (NWIS) of the United States Geological Survey (USGS) \citep{usgs}. The runoff station (USGS 04234000 Fall Creek) is located at [42°27'12"N, 76°28'22"W] and has a continuous record from February 1925 to the current year. Rainfall data is collected from the Global Historical Climate Network-Daily (GHCN-D) database that integrates station-based observations from numerous sources in producing daily climate records with quality assurance \citep{durre2008strategies,durre2010comprehensive,menne2012overview}. The rainfall station (GHCND:USC00304174) is maintained by Cornell [42°26’57"N, 76°26’57"W] and has a continuous record from April 1925 to the current year. And data over 1925-2017 is used in the model validation part. 

After the model validation, the framework is also tested on other NY watersheds. We try to examine the spatial variability of rainfall-runoff responses in this part and therefore, collect data for many small watersheds but over a shorter study period. 102 watersheds are selected and their information can be found in the \href{https://github.com/cruiseryy/MCMC_supplementary/blob/main/ID_AREA_LOC.txt}{Supplementary Materials}. Daily runoff data is collected from the USGS with a temporal coverage of 1979-2006. Considering no rainfall stations can be found near some of the USGS stations, gridded rainfall is used as collected from the Climate Prediction Center (CPC) Unified Gauge-Based Analysis of Daily Precipitation over CONUS provided by the NOAA/OAR/ESRL PSL, Boulder, Colorado, USA, from their website at https://psl.noaa.gov/data/gridded/data.unified.daily.conus.html. Quality control is performed on gauge records from multiple sources over 30,000 stations to provide a suite of unified precipitation products in the CPC data project \citep{xie2007gauge,chen2008assessing,xie2010cpc}. The CPC gridded rainfall has a spatial resolution of 0.25 $^o$ longitude by 0.25 $^o$ latitude and for each USGS station, rainfall from the nearest grid cell is used. 

Since we want to focus on monitoring changes in local hydrological responses at a inter-annual timescale in this study, the predominant effects of seasonality are removed by limiting each 'batch' to only one boreal summer period (June-July-August). And instead of separating individual storm events before fitting the hydrology model, data over the whole summer period is used which essentially averages the parameter estimations over multiple storm events. Each summer period can be adaptively extended by moving the starting date earlier so that effects of prior storm events will not be missed. The extension is limited to [0, 10] days and stops when runoff at the starting date is smaller than some thresholds (e.g., the 50$^{th}$ percentile value of runoff over the whole summer period). A summer period is dropped if it includes missing values or has significantly high runoff rate at the starting date even after the extension procedure. This step is to guarantee consistency in lengths of summer periods and thus in comparing model performance across different periods. The main reason is because even at the same statistical significance levels, the absolute value of a model skill metric (e.g., the Pearson's correlation coefficient used in this study) can change when the sample size changes.

\section{Methods}

\subsection{Synthetic Data Generator}
A two-step synthetic data generator is developed here to 1) sample rainfall based on historical records and 2) to produce runoff with some given noise levels for a typical summer period (i.e., 92 days). The rainfall events are first assigned using a Markov chain that transitions between the rainy (R) and sunny (S) states. The transition probabilities are estimated based on the historical Fall Creek rainfall events during the selected summer periods and the transition matrix is shown in Table 1. The initial state follows a Bernoulli distribution estimated based on the rainfall data at the starting dates of the selected summer periods such that $P(R_0) = 0.49$ and $P(S_0) = 0.51$.

\begin{table}[h!]
    \caption{The transition matrix estimated using historical rainfall events during the selected summer periods for the Fall Creek watershed}
    \centering
    
    \begin{tabular}{c|c|c}
    \toprule
        Current $\backslash$ Previous & Sunny & Rainy \\
     \midrule
        Sunny & 0.66 & 0.51 \\
        Rainy & 0.34 & 0.49 \\
    \bottomrule
    \end{tabular}
    
    \label{tab1}
\end{table}

After generating a time series of rainfall 'indicators' that determine the rainy-sunny state on each day, the intensities are simply sampled from historical rainfall intensities with replacements (i.e., bootstrapping). Assume the local rainfall-runoff response is governed by some model denoted by $\mathcal{M}$ such that $\tilde{R} = \mathcal{M}(P;\Theta)$. $\tilde{R}$ is the resulting runoff and $\Theta$ is a set of hidden parameters that are determined by physical properties and modulate the hydrological response. Then the noisy synthetic runoff data ($R_t$) can be generated by 

\begin{equation}
    R_t = \mathcal{M}(P_t;\Theta_t) + \varepsilon_t
\end{equation}

A time-invariant additive Gaussian noise is assumed such that $\varepsilon_t \sim N(0,\sigma^2)$. This simplifies the comparison of model skill with different noise levels as well as computations in the Bayesian estimation model (which will be detailed in the section 3.2). We will explore other options of noises in future work but such assumption has already been widely used in previous studies \citep{bates2001markov,moradkhani2005dual,stedinger2008appraisal}. And the specific form of the hydrology model $\mathcal{M}$ used in this study will be described in section 3.3.

\subsection{Bayesian Estimation Model}

We denote observations of rainfall $P_t$ and runoff $R_t$ at time $t$ by $D_t$ such that $D_t = (P_t, R_t)$. Assuming the local rainfall-runoff response is controlled by some set of hidden parameters $\Theta$ that cannot be easily measured, the goal of this study is to infer $\Theta$ given some observations of $D$ (i.e., $P(\Theta|D)$). Since it is hard to directly simulate $P(\Theta|D)$), the Bayes' Theorem is adopted as given by

\begin{equation}
    P(\Theta|D) = \frac{P(D|\Theta) \cdot P(\Theta)}{P(D)}
\end{equation}

The target PDF (often referred to as the posterior PDF) $P(\Theta|D)$ can then be estimated by calculating the likelihood PDF $P(D|\Theta)$, the prior PDF $P(\Theta)$, and the evidence PDF $P(D)$. The evidence PDF $P(D)$ can be regarded as a rescaling factor to make the resulting posterior PDF sum to one. Using the example of a continuous random variable of $\Theta$, $P(D)$ can be solved by integrating over all possibilities of $\Theta$ as given by

\begin{equation}
    P(D) = \int P(D \cap \alpha) d \alpha = \int P(D|\alpha) \cdot P(\alpha) d\alpha
\end{equation}

Analytical solutions for the integral in Equation 3 are only available for some specific combinations of prior and likelihood functions \citep{welch1995introduction,eddy2004hidden}. To numerically estimate the integral is always computationally intensive and even impossible for high dimensional random variables. Instead of directly solving the Equation 2, we propose to simulate the posterior PDF using a MCMC approach \citep{gilks1995markov}, which avoid computing the evidence PDF. The classic Metropolis Hastings algorithm \citep{chib1995understanding} is adopted here and consists of 3 major steps for constructing a Markov process of which the stationary distribution ($\pi_\infty$) converges to the target posterior PDF.

\setlength{\parindent}{3ex} 
1. Choose a transition kernel (also sometimes referred to as a proposal distribution) $q(y|x)$ such that $q(y|x) > 0 $ for all $x,y \in \chi$ ($\chi$ is the parameter space).

2. Define an acceptance rate $\alpha_{x,y}$ as given by

\begin{equation}
    \alpha_{x,y} = \min \left( 1, \frac{\pi_\infty(y)q(x|y)}{\pi_\infty(x)q(y|x)} \right)
\end{equation}

3. Repeatedly simulate the Markov process: at time $k$, given state $x_k = x$

\setlength{\parindent}{6ex} 
i) simulate next state of the Markov process $y \sim q(y|x)$,\par
ii) generate $u$ from an uniform distribution $u \sim U[0,1]$,\par
iii) if $u<\alpha_{x,y}$, set $x_{k+1} = y$; otherwise, set $x_{k+1}=x$.

\setlength{\parindent}{0ex} 
If we substitute the stationary distributions in Equation 3 with the posterior distributions, the acceptance rate is then computed by

\begin{equation}
    \alpha_{x,y} = \min \left( 1, \frac{P(D|y)P(y)q(x|y)}{P(D|x)P(x)q(y|x)} \right)
\end{equation}

The transition kernel essentially determines how fast the Markov process explores unknown parameter space. A variety of transition kernels have been developed and tested for improving efficiency of the MCMC algorithm \citep{yang2013searching,thawornwattana2018designing}. However, since our intent is to test feasibility of the Bayesian estimation model in tracking changes in hydrological responses rather than develop the most efficient algorithm, based on the principle of indifference \citep{keynes1921chapter}, the commonly used uniform distribution is assumed for the transition kernel $q(y|x)$ such that $y \sim U[x - \Delta x, x + \Delta x ]$. Assuming we have no prior knowledge of the watersheds, a uniform distribution is also assumed for the prior distribution $P(\Theta)$ such that $\Theta \sim U[\Theta_{min}, \Theta_{max}]$. The final acceptance rate can then be computed by 

\begin{equation}
    \alpha_{x,y} = \min \left( 1, \frac{P(D|y)}{P(D|x)} \cdot \mathbf{1}_{[\Theta_{min},\Theta_{max}]}(y) \right)
\end{equation}

where $\mathbf{1}_A(y)$ is an indicator function which returns 1 if $y \in A$ and 0 if not. Compared to the posterior distribution $P(\Theta|D)$, the likelihood distribution $P(D|\Theta)$ is usually easier to estimate when the governing model is already known. However, the simulated Markov process can be trapped by local optima when using the uniform transition kernel with small step sizes \citep{tjelmeland2001mode}, and this effect is particularly significant when simulating high dimensional random variables. We propose to initialize the Markov process with a searching procedure over some coarse grids in the range of $[\Theta_{min}, \Theta_{max}]$. The size of a coarse grid is set to $\frac{\Theta_{max}-\Theta_{min}}{10}$ and thus the total number of grids is $10^p$ for a $p-$dimensional parameter variable. $P(D|\Theta)$ is estimated using center parameter values of each coarse grid cell and the parameter value that maximizes $P(D|\Theta)$ is used to initiate the Markov process.

\subsection{Instantaneous Unit Hydrograph Model}

The Instantaneous Unit Hydrograph (IUH) model is a widely used conceptual model for runoff estimation and has a long history of successful implementation in hydrology \citep{gupta1980representation,jakeman1990computation,lee1997geomorphology,grimaldi2012parsimonious}. And it is well-established that IUHs can reflect some physical properties of a watershed like geomorphology characteristics \citep{gupta1980representation,lee1997geomorphology,da1997use}. This fact suggests potential of monitoring changes in local rainfall-runoff responses by tracking changes of IUHs. To study dynamics of IUHs in a quantitative way, here we propose to use a parameterized form as given by

\begin{equation}
    h(t;\lambda,k,\theta) = \lambda \frac{1}{\Gamma(k)\theta^k}t^{k-1}e^{-t/\theta}
\end{equation}

The formula is essentially a rescaled Gamma distribution: $\lambda$ is a scaling factor that accounts for loss of water in runoff generation processes (e.g., evapotranspiration); and $(k,\theta)$ are shaping parameters for the Gamma distribution. The Gamma distribution is used for its flexibility in modeling different types of IUHs and capability of simulating the low-pass filtering effect \citep{kirchner2000fractal,bhunya2003simplified}. Substituting the parameterized IUH into $\mathcal{M}$ in Equation 1, we can approximate the noisy runoff by

\begin{equation}
    R(t) = \tilde R(t) + \varepsilon_t = \sum^T_0 P(t-\tau) h(\tau;\lambda,k,\theta) + \varepsilon_t
\end{equation}

where the resulting runoff $\tilde R$ is modeled as the convolution of rainfall and the IUH $(\tilde R(t) = P(t) \otimes h(t; \Theta))$. $T$ is set to 14 days in this study. The probability of some given runoff observations conditioned on a set of parameters $(\lambda,k,\theta)$ can then be easily modeled using a Gaussian distribution ($R(t) \sim N(\tilde R(t), \sigma^2)$) and the acceptance rate in Equation 6 can be computed by

\begin{equation}
    \alpha_{x,y} = \min \left( 1, \exp \left( \frac{-1}{2\sigma^2} \sum_{t=1}^{N} (R(t)- \tilde R(t;h(y)))^2 - \frac{-1}{2\sigma^2} \sum_{t=1}^{N} (R(t)- \tilde R(t;h(x)))^2 \right) \cdot \mathbf{1}_{[\Theta_{min}, \Theta_{max}]}(y) \right)
\end{equation}

The hyperparameter $\sigma^2$ can be approximated by the mean squared error (MSE) using the MLE parameter estimations \citep{stedinger2008appraisal} as given by

\begin{equation}
    \sigma^2 = \frac{\sum_{t=1}^N(R(t) - \tilde R^{MLE}(t) )^2}{N}
\end{equation}

A physical interpretation of Equation 10 is that if we assume observations are governed by the proposed model, then MSE is minimized using the MLE parameters and should represent the true noise level. In this study, we denote the mean values of parameter samples from MCMC by the Bayesian parameter estimations. We replace the MLE parameter estimations with the Bayesian parameter estimations and Equation 10 is implemented in an iterative way:

\setlength{\parindent}{3ex} 
1. The PDF of $P(\Theta|D)$ is simulated using MCMC with some arbitrarily assigned constant for $\sigma^2$.

2. $\tilde{R} ^ {MLE}$ is approximated using the Bayesian parameter estimations and $\sigma^2$ is updated.

3. The MCMC is repeated using updated $\sigma^2$ and step 2-3 are repeated.

\setlength{\parindent}{0ex}
For simplicity, only one iteration is done and a prelim analysis shows that one iteration can already guarantee convergence for the approximated $\sigma^2$ ($R^2 = 0.998$ for 65 samples in comparing the approximated $\sigma^2$s after one and two iterations).

\section{Results \& Discussions}

\subsection{Model Validation}
\subsubsection{Synthetic Experiment}

Our analytical framework is first tested using synthetic data of summer rainfall and runoff. For each summer period, the IUH parameters are assumed constant and are drawn from an uniform prior distribution with ranges shown in Table 2. These ranges are arbitrarily assigned to cover a broad parameter space and the same prior distribution is also used in the following analysis.

\begin{table}[h!]
	\caption{Ranges of IUH parameter values $[\Theta_{min}, \Theta_{max}]$}
	\centering
	\begin{tabular}{llll}
		\toprule
		Parameters     & $\lambda$     &  $k$ & $\theta$ \\
		\midrule
		Ranges & $[0, 0.6]$ & $[0, 6]$ & $[0, 10]$ \\
		\bottomrule
	\end{tabular}
	\label{tab:table}
\end{table}

Model skill is evaluated using relative errors which are defined as ratios of absolute errors of parameters and the corresponding ranges ($err(\Theta) = \Delta \Theta \slash (\Theta_{max} - \Theta_{min})$). And the model is tested with noises of different levels. To facilitate the comparison of synthetic experiments and case studies using observational data, we use a relative metric of the signal-to-noise ratio ($SNR$) here rather than the absolute noise amplitudes. And $SNR$ is simply defined as a ratio of the standard deviation of noisy runoff to the standard deviation of noise ($SNR = \sigma(R) \slash \sigma (\varepsilon)$). Figure 1 shows model skill as a function of $SNR$ for three IUH parameters $(\lambda, k, \theta)$. At each $SNR$, the synthetic experiment is repeated for 50 times. The $25^{th}-75^{th}$ percentile ranges are plotted as shaded areas and the median model skill are plotted in diamond lines. A comparison between model skill of the initial grid searching (blue) and the Bayesian estimation model (red) demonstrates that the MCMC can always improve accuracy of parameter estimations when $SNR$ is greater than 1. And when the noise is comparable to even larger than the signal (i.e., $SNR < 1$), performance of both method degrades significantly in terms of median relative errors as well as uncertainty ranges. A reference $SNR$ level is approximated using the ratio of standard deviation of noisy runoff observations to that of runoff residuals. The runoff residual is calculated by fitting the IUH model with Bayesian parameter estimations using the observational data from the Fall Creek watershed. The smallest value of $SNR$ of all selected summer periods is also plotted in Figure 1 and the results indicate a good potential performance when applying our Bayesian estimation model on real-world observational data. For any $SNR$ greater than the reference level, the median relative errors should be smaller than 6\% for all parameter estimations (or smaller than 3\% for ($\lambda, k$)).

\begin{figure}[h!]
    \centering
    \includegraphics[width=140mm]{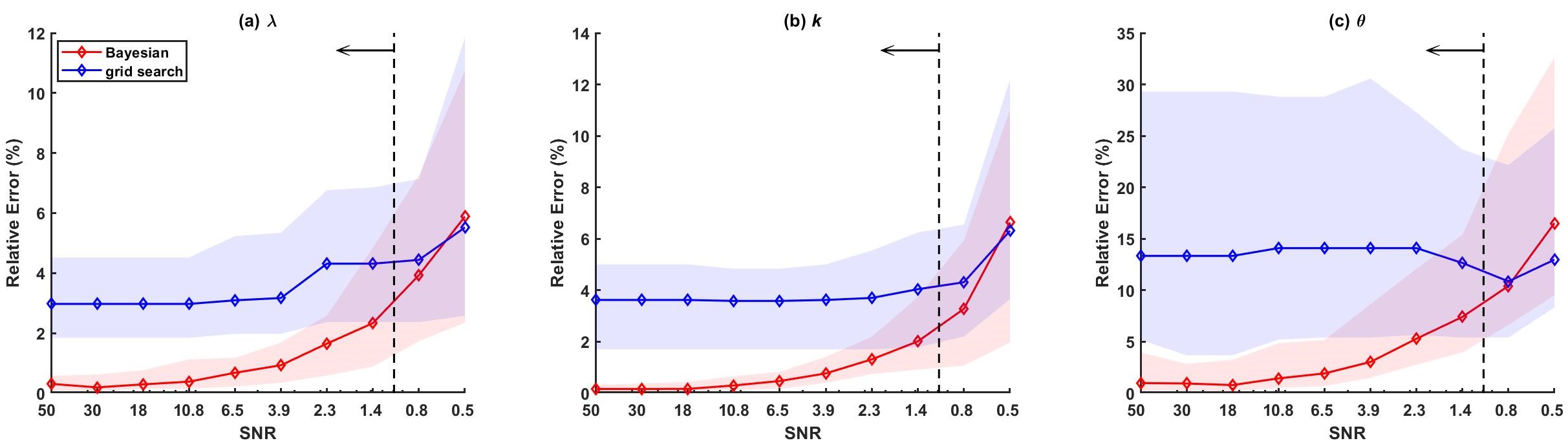}
    \caption{Model skill measured by relative errors as a function of $SNR$ for (a) $\lambda$, (b) $k$, and (c) $\theta$. Performance of the initial grid searching and the Bayesian estimation model are plotted in blue and red lines, respectively. At each $SNR$, the synthetic experiment is repeated for 50 times. The $25^{th}-75^{th}$ percentile ranges are plotted as shaded areas and the median model skill are plotted in diamond lines. The reference $SNR$ based on observational data from the Fall Creek watershed in indicated by a vertical black dashed line. And observational data from the Fall Creek watershed have $SNR$ values no smaller than this reference level for all selected summer periods as indicated by the leftward arrow.}
    \label{fig1}
\end{figure}

\subsubsection{Validation Using Observational Data}

In addition to using synthetic data, we also examine the model performance using observational data of rainfall and runoff from the Fall Creek watershed. 65 summer periods are selected over the whole period of 1925-2017 and IUH parameters are estimated for each summer period. Since true parameter values are not known here, MLE parameter estimations are used in the comparison. The MLE parameter estimations are computed by randomly searching in the parameter space $[\Theta_{min}, \Theta_{max}]$ to minimize MSE between the observed and estimated runoff and should represent the best possible parameter values given our conceptual IUH model. 

\begin{figure}[h!]
    \centering
    \includegraphics[width=140mm]{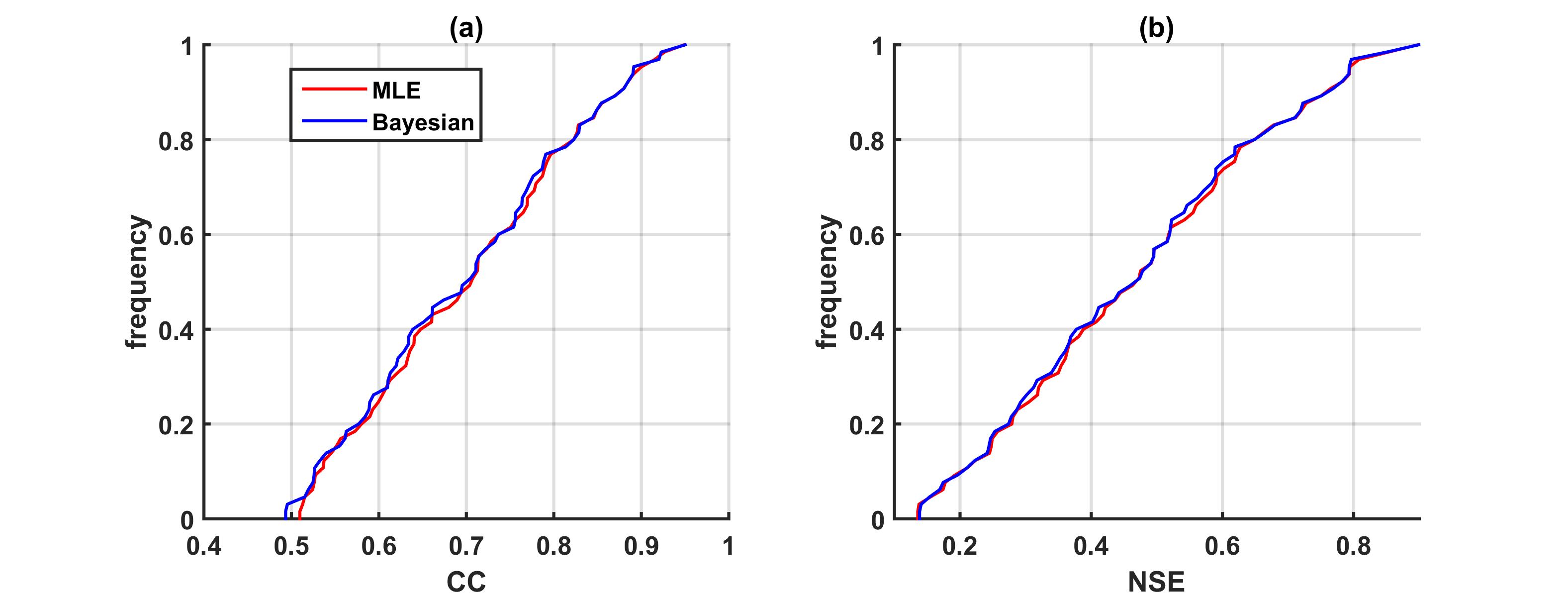}
    \caption{Empirical cumulative distribution functions of model skill measured by (a) Pearson's correlation coefficients (CC) and (b) Nash-Sutcliffe Efficiency (NSE) scores for runoff estimations. Performance of models fitted using MLE and Bayesian parameter estimations are plotted in the red and blue curves, respectively.}
    \label{fig2}
\end{figure}

We first fit the IUH models using both MLE and Bayesian parameter estimations and their performance in estimating runoff are compared in Figure 2. Very negligible difference are observed in model skills using MLE and Bayesian parameters as measured by both CCs and NSE scores. And comparable good model skill are observed for runoff estimations using both sets of parameters: the median CCs are greater than 0.7 and the median NSE scores are greater than 0.49. The good model skill should also validate the Gamma distribution-based IUH in modeling the rainfall-runoff response in the Fall Creek watershed. No statistical significance levels are reported in the figure since the effective degrees of freedom depend on autocorrelations in the data \citep{afyouni2019effective} and vary significantly across different summer periods (considering autocorrelations in runoff/rainfall are largely affected by frequencies and intensities of storm events.)

\begin{figure}[h!]
    \centering
    \includegraphics[width=140mm]{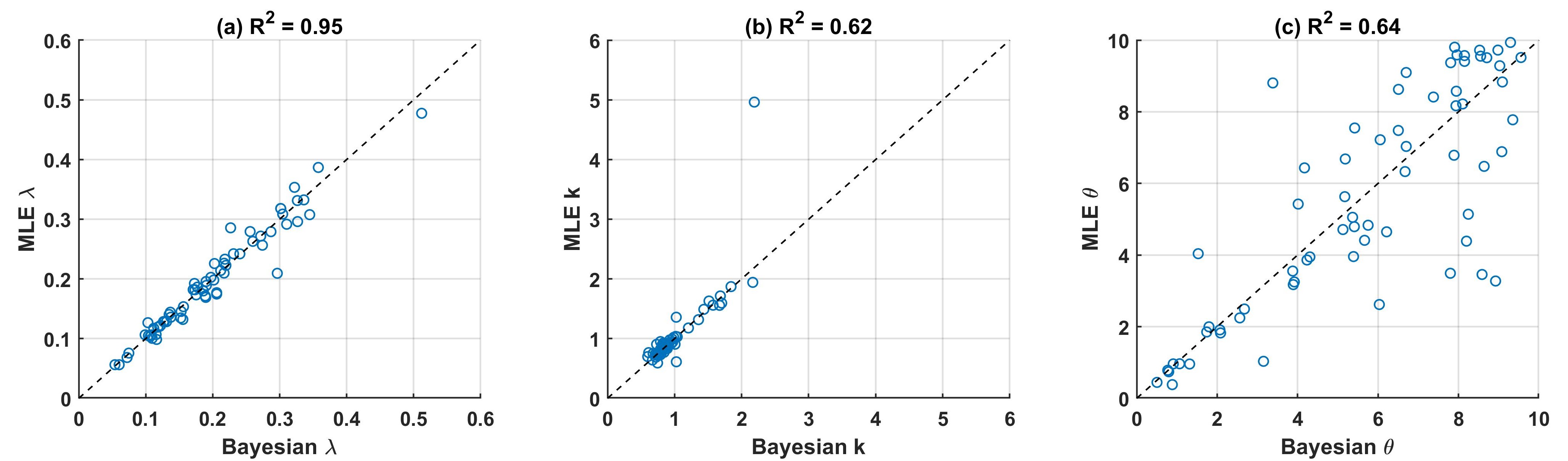}
    \caption{Comparison between the Bayesian (x axis) and MLE (y axis) parameter estimations for (a) $\lambda$, (b) $k$, and (c) $\theta$. The 1:1 reference lines are plotted in black dashed lines and the coefficients of determination ($R^2$) for 65 summer periods are reported in the subplot titles.}
    \label{fig3}
\end{figure}

The Bayesian parameter estimations are then compared against their MLE counterparts as shown in Figure 3. Great consistency are observed between the MLE and Bayesian parameter estimations: the $R^2$ values calculated using the 1:1 reference line are 0.95, 0.62, and 0.64 for $(\lambda, k, \theta)$, respectively. Of the three parameters, the highest $R^2$ value is found for $\lambda$. A possible explanation can be that MSE in calculating $\alpha_{x,y}$ (in Equation 9) is more 'sensitive' to $\lambda$ since it solely determines the integral amplitude of the IUH while the shape parameters $(k, \theta)$ only collectively determine the IUH shape. For example, different combinations of $(k,\theta)$ can result in very similar IUH shapes and therefore, more deviations between the MLE and Bayesian estimations are expected for $(k,\theta)$.

Also, the Bayesian estimations of $\lambda$ are found to be consistent with the runoff coefficients approximated using the ratio of total amount of runoff to that of rainfall. The $R^2$ value calculated using the 1:1 reference line is 0.65 and the correlation coefficient is 0.87 for 65 summer periods (one tailed p-value $<$ 0.001). On one hand, this should should validate the Gamma distribution-based IUH in simulating the local rainfall-runoff response since $\lambda$ is specifically designed to account for the loss of water. On the other hand, the approximated runoff coefficients can be used as the true $\lambda$ values and thus the results suggest that the Bayesian estimation model can well infer and track the time-varying hidden parameters. 

\subsection{Model Application}
\subsubsection{A Historical Change Point in the Fall Creek Watershed}
The analytical framework is then used to study dynamics of rainfall-runoff responses in the Fall Creek watershed through 1925-2017. Since it is hard to compare IUHs of different shapes in a quantitative way, we first define a metric of the initial decay rate ($IDR$) as given by 

\begin{equation}
    IDR = \frac{\partial h \slash \partial t}{h}|_{t = 0.05} = \left( \frac{k-1}{t} - \frac{1}{\theta} \right) |_{t = 0.05}
\end{equation}

$IDR$ is essentially the relative initial derivative of IUH and is estimated at $t = 0.05$ [day] to avoid a zero denominator. Using $IDR$, we can easily distinguish two typical IUHs: the IUHs that monotonically decreases over time are defined as the 'diffusion' type while those with wave-like patterns are defined as the 'advection' type. A diagram of the two IUH types is shown in Figure 4 and an IUH with greater-than-zero $IDR$ falls into the category of 'advection' and 'diffusion' if otherwise. The two types can have direct use in informing flooding mitigation since they determine timings and rates of peak flow differently. Also the two IUH types can reflect different water travel times and therefore indicate different hydrological properties.

\begin{figure}[h!]
    \centering
    \includegraphics[width=80mm]{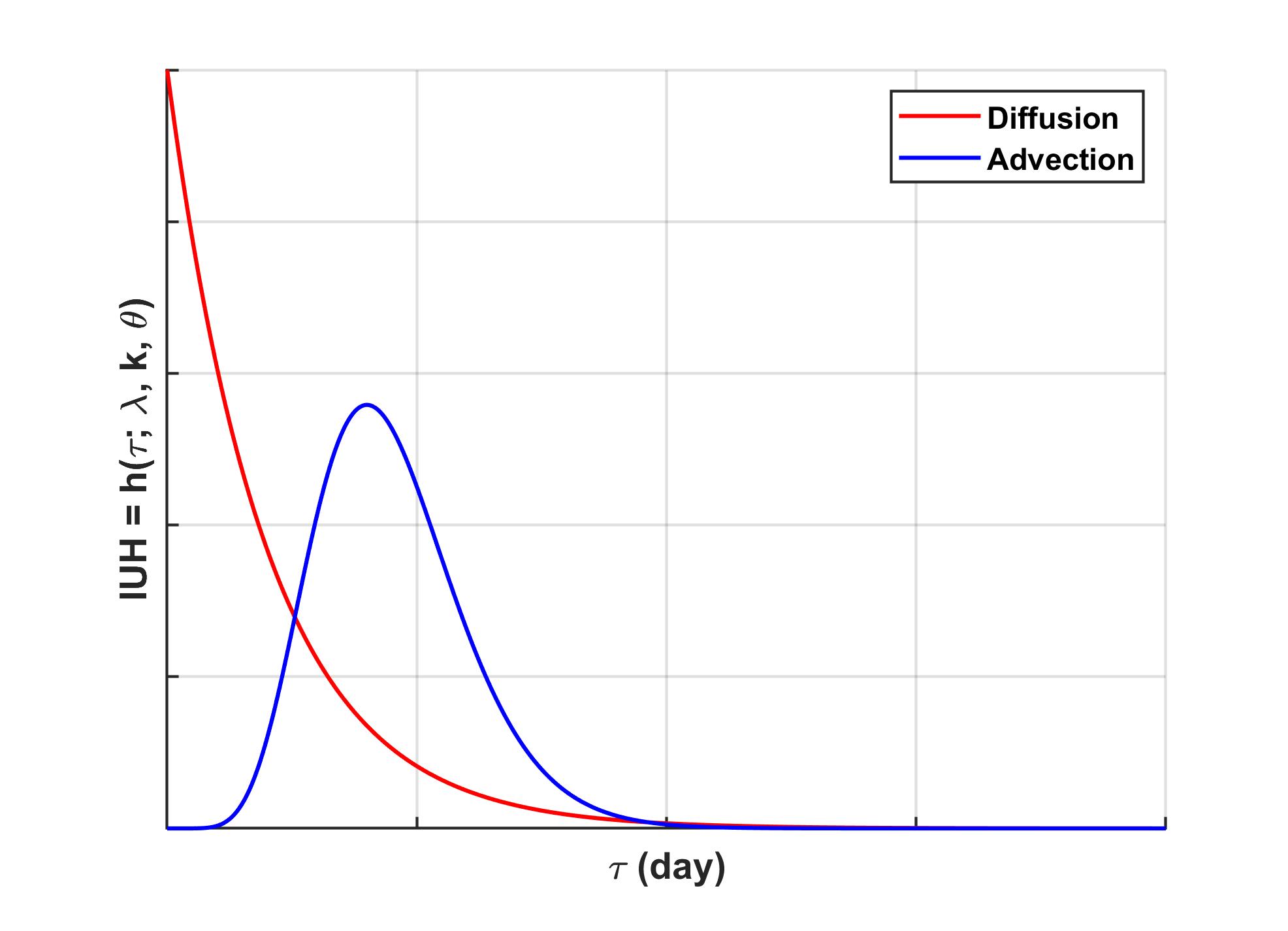}
    \caption{A diagram of two types of IUHs: the 'diffusion' type (red) and the 'advection' type (blue).}
    \label{fig4}
\end{figure}

\begin{figure}[h!]
    \centering
    \includegraphics[width=140mm]{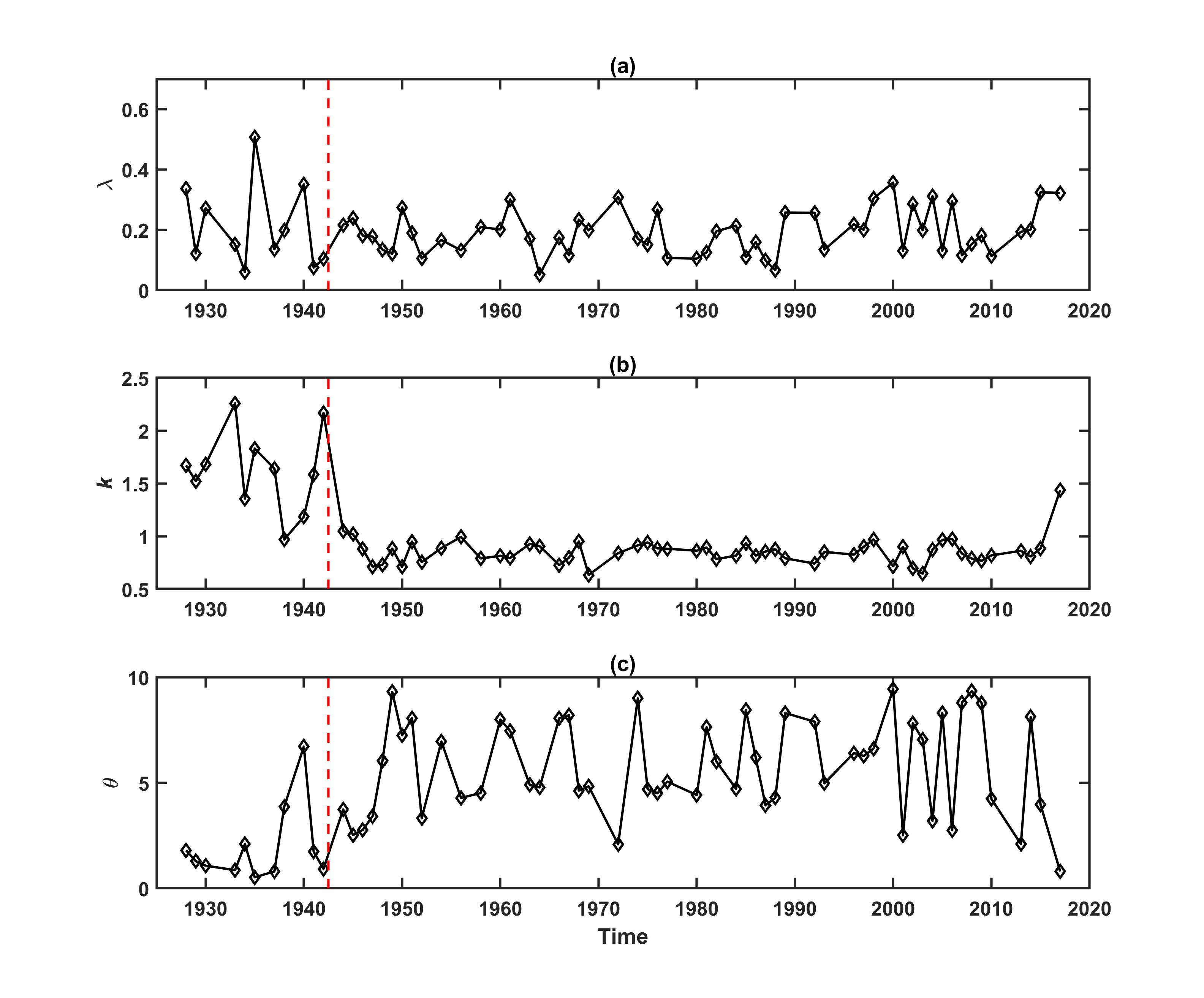}
    \caption{Time series of Bayesian estimations of (a) $\lambda$, (b) $k$, and (c) $\theta$. The year of 1943 when Cornell moved its rainfall station is indicated by a vertical red dashed line. }
    \label{fig5}
\end{figure}

\begin{figure}[h!]
    \centering
    \includegraphics[width=140mm]{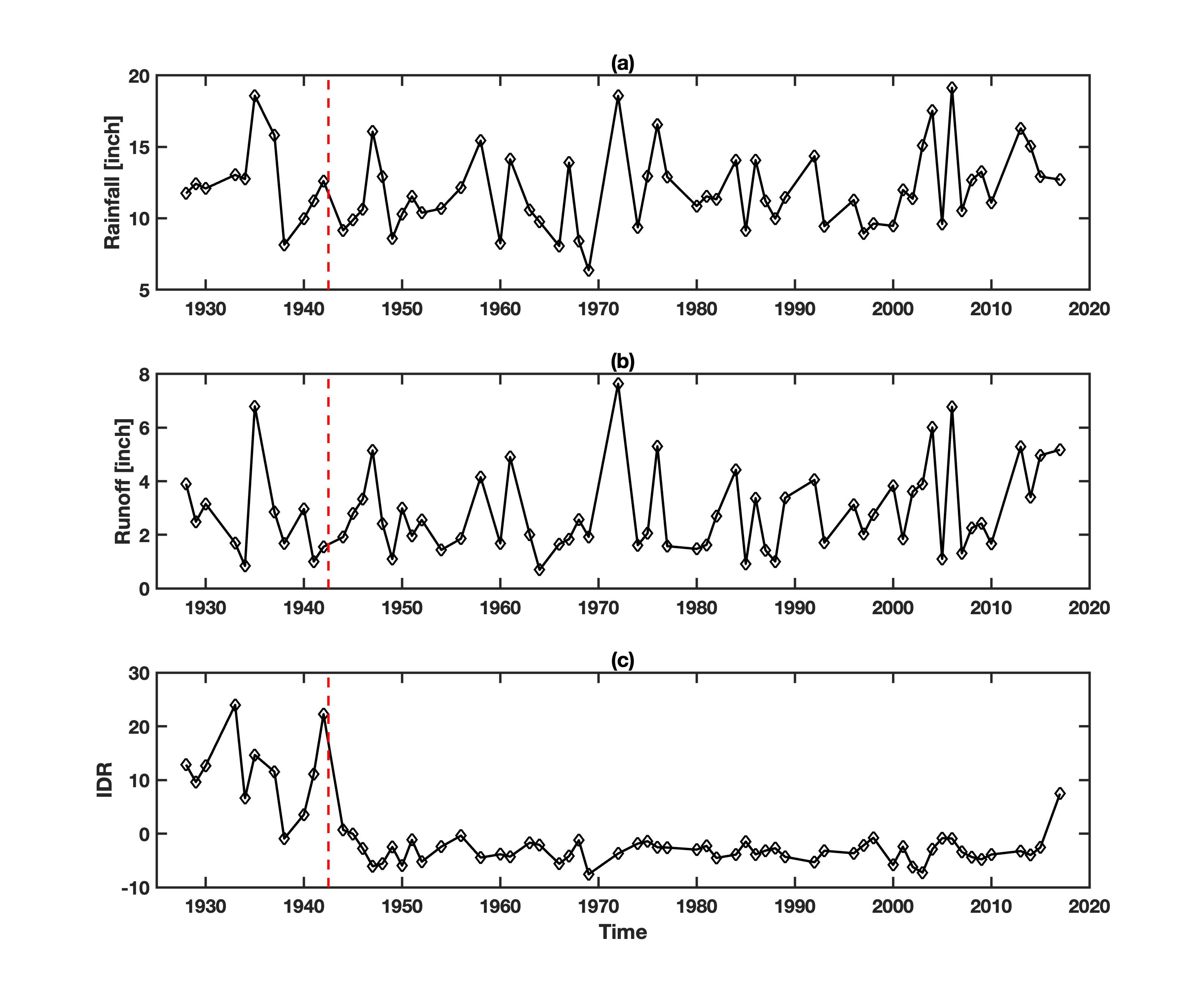}
    \caption{Time series of (a) total summer rainfall, (b) total summer runoff, and (c) $IDR$. The year of 1943 when Cornell moved its rainfall station is indicated by a vertical red dashed line. }
    \label{fig6}
\end{figure}

Time series of Bayesian parameter estimations are plotted in Figure 5. Systematic shifts are observed are observed around 1943 for the shape parameter $(k,\theta)$: values of $k$ drop significantly and values of $\theta$ increase by a smaller margin after around 1943. No such change point is found for the rescaling parameter $\lambda$. The decreasing $k$ can turn IUHs from the 'advection' type to the 'diffusion' type while the increasing $\theta$ has the opposite effect. We find that historical patterns of IUHs are dominated by changes in $k$ when we compare time series of $IDR$ with those of summer rainfall and runoff in Figure 6. Before 1943, the local IUH was dominated by the 'advection' type and after 1943, the local IUH switched to the 'diffusion' type, which is consistent with the rapid drop of $k$ values. However, such change cannot be learned by looking at time series of summer rainfall and runoff. And this suggests that the change could occur only in the temporal distribution of water. Physically, this change point can indicate that water in the Fall Creek watershed flows faster (i.e., has shorter travel time) after 1943, which can possibly be due to changed in land uses like construction of paved roads. However, we could not find any strong support evidence from chronicles of the Fall Creek watershed or the Ithaca city so far. Interestingly, the change point coincided with relocation of the rainfall station when Cornell moved it from the Roberts Hall (on the ground outside of the building) [42°26’55"N, 76°28’45"W] to the College of Agriculture Experimental Farm at the Caldwell Field [42°26’56"N, 76°27’38"W] in June 1943. While the relocation of rainfall station can lead to changes in latency between rainfall and runoff observations, we expect the effect to be minor since 1) the rainfall stations was only move one mile east in 1943 (for reference, a hurricane typically travels at a forward speed of 5-20 miles per hour \citep{king1978radar}) and 2) the station was moved again to its current location on the Game Farm Road [42°26’57"N, 76°26’57"W] in June 1969. We did not observe any changes in IUH parameters associated with the second relocation. But still, the Bayesian estimation model manages to identify a systematic shift in the local rainfall-runoff response that cannot be learned by looking at the easy-to-measure variables like rainfall or runoff.

\subsubsection{Relationships between IUHs and Watershed Sizes}
In the previous section, the Bayesian estimation model is applied to examine temporal variability of rainfall-runoff responses. In this section, we extend its application to examining rainfall-runoff responses of many watersheds by repeating the analysis for 102 NY watersheds. Considering a relatively short study period is used (1979-2006) here, the hydrological responses are assumed constant and the median $IDR$s for each watershed are compared to investigate the spatial patterns. A summer period is only selected when the NSE score is better than 0.2 for the runoff estimation using the Bayesian parameters. And only watershed with more than 5 selected summer periods are considered. This is because the IUH cannot represent the local rainfall-runoff response when it fails at estimating the runoff.

Figure 7a shows a map of median $IDR$s for all selected watersheds (or watershed with 'good' NSE scores). While more positive $IDR$ values are found in the northwestern region, this pattern can be due to a non-uniform distribution of watersheds of different sizes. To examine relationships between rainfall-runoff responses and watershed sizes, we first define a watershed length scale simply using the square root of the watershed drainage area. A scatter plot between the median $IDR$s are watershed length scales is shown in Figure 7b and a positive relationship is observed. A smaller watershed is more likely to be governed by the 'diffusion' type of IUH ($IDR<0$) while a larger watershed is more likely to be governed by the 'advection' type of IUH ($IDR>0$). The positive relationship is found to be statistically significant as examined by a F test (one-tailed p-value $<$ 0.01). The results here appear to be intuitive since water usually needs longer travel time to reach the outlet in a larger watershed. But the underlying physical causalities require further rigorous investigation with runoff generation processes and channel network properties taken into consideration.

\begin{figure}[h!]
    \centering
    \includegraphics[width=140mm]{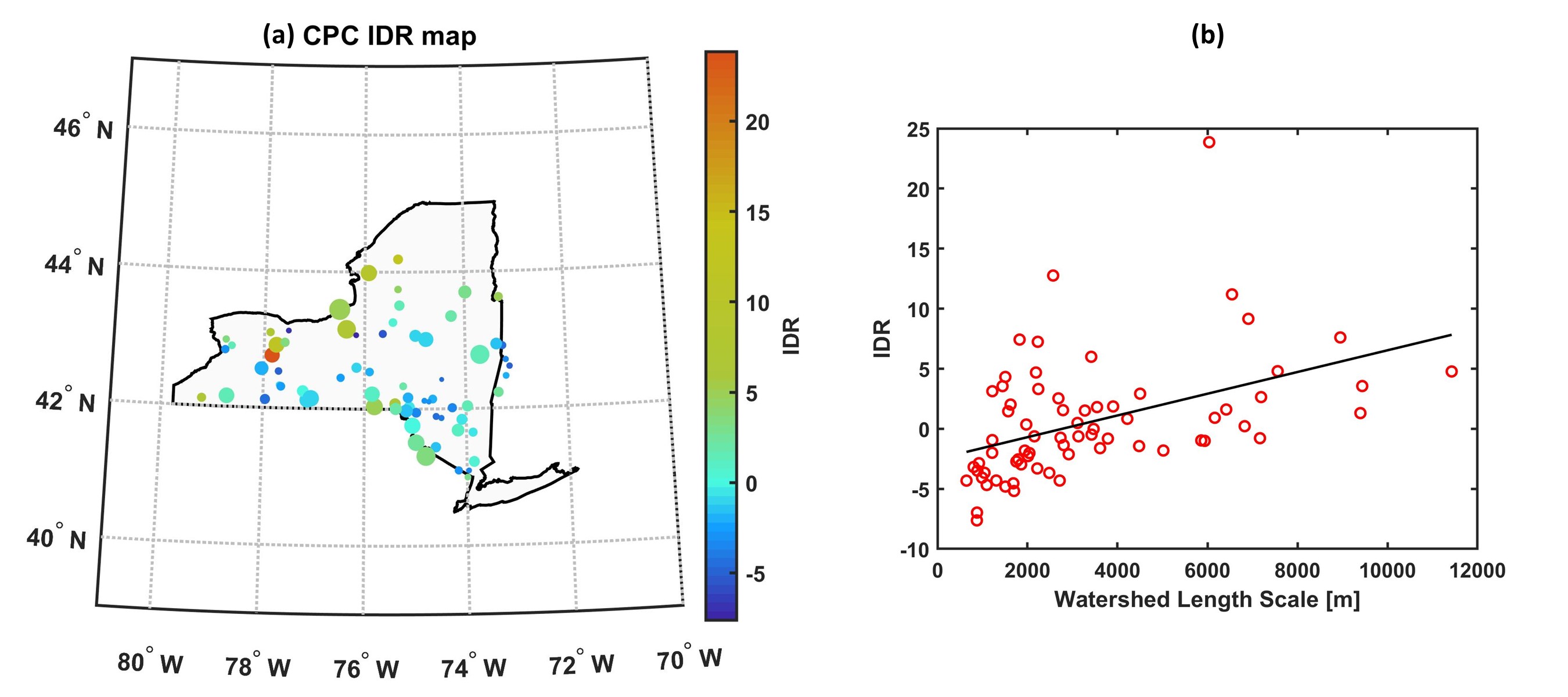}
    \caption{(a) A map of median $IDR$s for watersheds with 'good' NSE scores where sizes of circles are proportional to watershed length scales. (b) A scatter plot between watershed length scales and median $IDR$s and its fitted line (black).}
    \label{fig7}
\end{figure}

\section{Conclusion}
In this study, performance of a Bayesian estimation model (i.e., MCMC) is examined in inferring hydrology model parameters given observations of rainfall and runoff for small watersheds. The IUH conceptual model based on a modified Gamma distribution is adopted and the whole analytical framework is validated using both synthetic and observational data. In the synthetic simulation, great accuracy is observed for Bayesian parameter estimations with relative errors smaller than 6\% when the noise is smaller than the reference level. In absence of true parameter values, the Bayesian parameter estimations are compared against the MLE parameter estimations when observational data is used. Comparable model skill are observed for runoff estimations using MLE and Bayesian parameter estimations and good consistency is observed for all parameters between the two sets. Then the Bayesian model is applied to study temporal and spatial variability of rainfall-runoff responses by tracking parameter estimations over time or different watersheds. A systematic shift is identified for the Fall Creek watershed when its IUH switched from the 'advection' type to the 'diffusion' type around 1943. This change point cannot be learned by looking at the easy-to-measure variables like summer rainfall and runoff. The underlying cause remains unknown considering the shift can be due to either or both effects of changes in land uses and the relocation of the rainfall station. Furthermore, rainfall-runoff responses of 102 NY watersheds are studied but over a shorter study period. A statistically significant positive relationship is observed between $IDR$s and watershed sizes, which suggests that smaller watersheds are more likely to be governed by IUHs of the 'diffusion' type. Overall, our study demonstrates feasibility of the Bayesian estimation model in monitoring dynamics of hydrological responses and detecting change points for small watersheds by tracking the hidden variables (i.e., model parameters). And the model shows potential as a rapid surveillance tool to aid adapting local strategies of water resource management and disater mitigation to changes of rainfall-runoff responses by iteratively updating the analytical framework with new bathes of observational data. 

\section{Acknowledgement}

We want to thank Dr. Wilfried Brutsaert from the School of Civil and Environmental Engineering at Cornell University for his helpful comments on historical hydrology conditions of the Fall Creek watershed, William Coon from USGS for compiling historical records of the runoff gauge, and Keith Eggleston and Mark Wysocki from the Department of Earth and Atmospheric Sciences at Cornell University for compiling historical records of the rainfall station.

\bibliographystyle{unsrtnat}
\bibliography{references}  






\end{document}